\documentclass[aps,pra,twocolumn,superscriptaddress,nofootinbib,longbibliography]{revtex4-2}

\usepackage[T1]{fontenc}
\usepackage{lmodern}
\usepackage{amsmath,amssymb,amsthm,bm,bbm}
\usepackage{graphicx}
\usepackage{xcolor}
\usepackage[colorlinks=true,citecolor=blue,linkcolor=blue,urlcolor=blue]{hyperref}
\hypersetup{pdftitle={Fer expansion of driven spins: closed-form recursion and no-go theorems for termination},
  pdfauthor={Lorenzo Bagnasacco, Daniel Burgarth, Vittorio Giovannetti, Kazuya Yuasa}}

\theoremstyle{plain}
\newtheorem{theorem}{Theorem}
\theoremstyle{remark}
\newtheorem*{remark}{Remark}

\newcommand{\rmd}{\mathrm{d}}
\newcommand{\rmi}{\mathrm{i}}
\newcommand{\rme}{\mathrm{e}}
\renewcommand{\openone}{\mathbbm{1}}
\DeclareMathOperator{\ad}{ad}
\numberwithin{equation}{section}

\begin{document}
\title{Fer expansion of driven spins: closed-form recursion and no-go
theorems for termination}

\author{Lorenzo Bagnasacco}
\email{lorenzo.bagnasacco@sns.it}
\affiliation{NEST, Scuola Normale Superiore, Piazza dei Cavalieri 7, I-56126 Pisa, Italy}
\author{Daniel Burgarth}
\affiliation{Department Physik, Friedrich-Alexander-Universit\"at Erlangen--N\"urnberg, 91058 Erlangen, Germany}
\author{Vittorio Giovannetti}
\affiliation{NEST, Scuola Normale Superiore and Istituto Nanoscienze-CNR, Piazza dei Cavalieri 7, I-56126 Pisa, Italy}
\author{Kazuya Yuasa}
\affiliation{Department of Physics, Waseda University, Tokyo 169-8555, Japan}

\begin{abstract}
The Fer expansion factorizes the propagator of a driven quantum system
into exponentials of time-integrated effective Hamiltonians, which govern the
evolution left over by the previous factors. Every truncation is unitary, and
adding a factor squares the size of the remainder, but each effective
Hamiltonian is an infinite series of nested commutators of the previous one.
For $\mathfrak{su}(2)$, in every representation, from qubits to spins of any
dimension, we represent the effective Hamiltonians by time-dependent vectors
in $\mathbb{R}^3$ and resum this series exactly. The result is a closed map
from one vector to the next, in which the infinite series reduces to the
previous vector crossed once and twice with its time integral, weighted by
elementary functions. Using this map, we address the question of when the
expansion can be stopped after finitely many factors. This is possible
whenever the leftover evolution is known in closed form, for instance when
the effective Hamiltonian commutes with itself at all times or precesses
about a fixed axis with constant angular velocity. We prove that, for
real-analytic drives, neither happens beyond the first factor unless the
first exponential is already exact. Stopping the expansion therefore yields
no new exactly solvable drives.
\end{abstract}

\maketitle

\section{Introduction}
\label{sec:intro}

A spin driven along a time-dependent axis is the working model of
magnetic resonance~\cite{bloch} and, for qubits, of most quantum-computing
platforms~\cite{Vandersypen2005}. Its Hamiltonian is a time-dependent
combination of the generators of $\mathfrak{su}(2)$, up to a multiple of the
identity that only contributes a global phase. Hamiltonians of the same
form, in other representations of the algebra, describe a spin of any
size, an ensemble of two-level atoms driven collectively, and, through
Schwinger's oscillator model~\cite{sakurai}, two bosonic modes coupled by a
time-dependent beam splitter. In these settings, designing pulse sequences,
decoupling schemes, and gates requires the evolution operator $U(t)$
generated by the Hamiltonian $H(t)$, or an effective Hamiltonian that
approximates it~\cite{haeberlen1968,viola1999,koch2022}. Formally, $U(t)$ is given by
the time-ordered exponential $\mathcal{T}\exp[-\rmi\int_0^t\rmd s\,H(s)]$,
whose expansion in powers of $H$ is the Dyson
series~\cite{dyson1949radiation,dyson}. This is a formal solution only, and
truncating it breaks unitarity at every finite
order~\cite{blanes2009}.

\begin{figure}[t]
\centering
\includegraphics[width=\columnwidth]{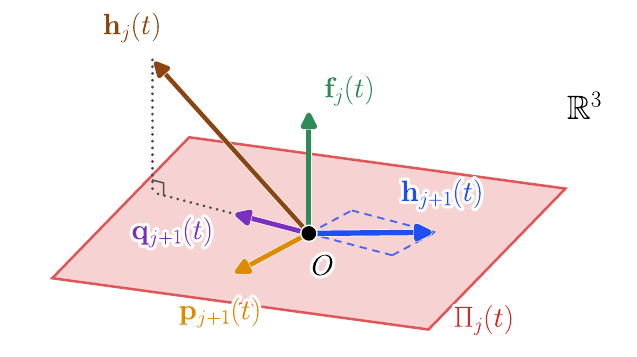}
\caption{One step of the Fer map on $\mathfrak{su}(2)$,
Eq.~\eqref{eq:su2-fer-map-closed}, drawn in the space $\mathbb{R}^3$ of the
coefficient vectors. The integrated drive
$\mathbf{f}_j(t)$, in green, is normal to the plane $\Pi_j(t)$, in red,
through the origin $O$. The
auxiliary vectors $\mathbf{p}_{j+1}(t)=\mathbf{f}_j(t)\times\mathbf{h}_j(t)$, in
orange, and $\mathbf{q}_{j+1}(t)=\mathbf{p}_{j+1}(t)\times\mathbf{f}_j(t)$, in
purple, span $\Pi_j(t)$; the latter lies along the projection of the drive
$\mathbf{h}_j(t)$, in brown, onto the plane (dotted). The output
$\mathbf{h}_{j+1}(t)=\alpha(\mathrm{f}_j(t))\,\mathbf{p}_{j+1}(t)
+\beta(\mathrm{f}_j(t))\,\mathbf{q}_{j+1}(t)$, in blue, is assembled from them
(dashed parallelogram) with the weights $\alpha$ and $\beta$ of
Eqs.~\eqref{eq:alpha-def} and~\eqref{eq:beta-def}, both evaluated at the norm
$\mathrm{f}_j(t)=\|\mathbf{f}_j(t)\|$, and never leaves $\Pi_j(t)$.
}
\label{fig:fer-step}
\end{figure}

Several methods avoid this problem by keeping the propagator unitary at
every order. The best known is the Magnus expansion~\cite{magnus,blanes2009,klarsfeld1989,iserles1999}, which writes the
propagator as a single exponential, $U(t)=\rme^{\Omega(t)}$, with $\Omega(t)$
skew-Hermitian and built from nested commutators of $H(t)$ at different
times. Its convergence and error estimates are well
charted~\cite{blanes2009,casas2007,moan2006}, where $\Omega(t)$ is an infinite series, in which each
new term adds one more commutator of the true Hamiltonian.

A second route exploits the algebraic structure of $H(t)$ instead of
expanding in nested commutators. If $H(t)=\sum_{a=1}^{\dim\mathfrak{g}}
u_a(t)\,\mathrm{T}_a$, with real coefficients $u_a(t)$, is built from
Hermitian generators $\mathrm{T}_a$ that close under commutation into a
finite-dimensional Lie algebra $\mathfrak{g}$, the Wei--Norman
theorem~\cite{weinorman,weinorman1964,wilcox} factorizes the propagator over
a basis of $\mathfrak{g}$, locally in time, into the ordered product
$U(t)=\prod_{a=1}^{\dim\mathfrak{g}}\rme^{-\rmi\xi_a(t)\mathrm{T}_a}$, and
turns the Schr\"odinger equation into a closed system of scalar ordinary
differential equations (ODEs) for the $\xi_a(t)$. Except for special cases
such as solvable algebras~\cite{weinorman1964}, the $\xi_a(t)$ can diverge at
finite times~\cite{guerrero2020}, beyond which the factorization has to be
restarted. The method has been revisited in tutorial form for Gaussian and
non-Gaussian quantum-optical models~\cite{qvarfort}, underlies unitary
integration~\cite{rau-unitary}, and its equations and their integrability
have been studied in detail~\cite{charzynski2013,altafini2005}; the same
factorizations are behind the Lie-group integrators of numerical
analysis~\cite{iserles2000}.

The Fer expansion~\cite{fer,blanes1998,bagnasacco2026fer} lies between
these two approaches and has been explored much less, although it has been
used to build numerical integrators~\cite{zanna1999} and compared with the
Magnus and Floquet expansions in solid-state NMR~\cite{takegoshi2015}. Like
Magnus, it is built from nested commutators alone and needs no prior
knowledge of the algebra generated by $H(t)$; like Wei--Norman, it writes the
propagator as a product of exponentials. The factors are peeled off one at a
time, each generated by the time integral of an effective Hamiltonian
$H_j(t)$ that governs the evolution left over by the previous ones. Every
truncation is therefore unitary, and what it leaves out is not an
uncontrolled error but the evolution generated by the next effective
Hamiltonian $H_{j+1}(t)$. Each factor also squares the size of this
remainder, so that, within the convergence domain of the
expansion~\cite{blanes1998}, recently enlarged by about
$30\%$~\cite{bagnasacco2026fer}, the error decreases doubly exponentially
with the number of factors~\cite{blanes2009}. The price is that each effective
Hamiltonian is an infinite series of nested commutators of the previous one,
which in general has to be truncated or evaluated numerically at every
order.

First we show that, for $\mathfrak{su}(2)$, the Fer expansion can
nevertheless be made explicit at every order: the infinite series of nested
commutators that defines each effective Hamiltonian can be summed exactly, in
every representation of the algebra. Each effective Hamiltonian is a
combination of the three generators of $\mathfrak{su}(2)$ with
time-dependent coefficients that form a vector in $\mathbb{R}^3$, and
commutators become cross products of these vectors. One step of the
recursion thus becomes an explicit map from each vector to the next: the new
vector is the previous one crossed once and twice with its time integral,
and the two terms are weighted by elementary trigonometric functions. Every
factor of the Fer product is then an explicit rotation of the spin.

We then use the closed map to answer a question that the structure of
the Fer product raises: can the product
stop by itself? If at some step the effective Hamiltonian commutes with
itself at all times, the next one vanishes identically, there is nothing
left to peel off, and finitely many factors give the exact propagator. More
generally, the product can be stopped as soon as the leftover evolution is
known in closed form, for instance when the effective Hamiltonian precesses
about a fixed axis with constant angular velocity, as in Rabi's rotating
field~\cite{rabi}. If a generic drive could be steered onto one of these two
classes after finitely many steps, the Fer expansion would produce new
exactly solvable drives, starting from the drive rather than from the
solution.

Such drives would be valuable, since closed-form propagators are rare.
For a qubit, the classic examples are Rabi's rotating field~\cite{rabi} and
the Landau--Zener sweep~\cite{Zener1932}. Larger families have been obtained
by reverse-engineering the two-level Schr\"odinger
equation~\cite{Barnes2012,Messina2014}, or through the dynamical invariants
of Lewis and Riesenfeld~\cite{lewis1969}, which trade the linear problem for
an auxiliary nonlinear equation, such as the Ermakov--Pinney equation of the
time-dependent harmonic oscillator~\cite{pinney1950}, and underpin the inverse
engineering of shortcuts to adiabaticity~\cite{GueryOdelin2019}. In every
case the drive is built from the solution, and not the other way around: for
a drive given in advance, the propagator is in general not known in closed
form.

Among the drives with a closed-form propagator, we consider two classes
on which the Fer product can be stopped: effective Hamiltonians that commute
with themselves at all times, and those that precess about a fixed axis with
constant angular velocity, as in Rabi's rotating field. Whether the product
can reach either class at some order is not obvious in advance, since the
recursion mixes the components of the drive nonlinearly at every step. We
prove two no-go theorems showing that it cannot, except trivially. More
precisely, for a real-analytic drive, an
effective Hamiltonian beyond the first falls into either class only if it
vanishes identically and the original drive already commutes with itself at
all times, in which case the first exponential is already exact: the
recursion creates no stopping point that was not already present in the
original drive.

The paper is organized as follows. Section~\ref{sec:fer} recalls the Fer recursion, and
Sec.~\ref{sec:framework} isolates what in it does not depend on the algebra,
including the condition under which the product stops.
Section~\ref{sec:su2} derives the closed map and shows how it makes every
order of the expansion explicit, Sec.~\ref{sec:su2-rigidity} proves the two
no-go theorems, and Sec.~\ref{sec:conclusion} discusses where these results
apply. The
derivations are collected in
Appendixes~\ref{app:conjugation}--\ref{app:derivative}.

\section{The Fer expansion}
\label{sec:fer}

Throughout, we set $\hbar=1$ and a dot denotes the total time derivative
$\rmd/\rmd t$. The evolution operator of a Hamiltonian $H(t)$ obeys
\begin{equation}
    \frac{\rmd}{\rmd t}\,U(t) = -\rmi H(t)\,U(t)\;, \qquad U(0)=\openone\;.
    \label{eq:schrodinger}
\end{equation}
Following Refs.~\cite{fer,blanes1998}, and in the spirit of the
average-Hamiltonian and Floquet--Magnus methods of magnetic
resonance~\cite{haeberlen1968,mananga2011} and of dynamical
decoupling~\cite{viola1999}, one writes it as an ordered product of
exponentials,
\begin{equation}
    U(t) = \rme^{F_1(t)}\rme^{F_2(t)}\cdots\rme^{F_j(t)}\,U_{j+1}(t)\;,
    \label{eq:fer-product}
\end{equation}
in which the remainder obeys a Schr\"odinger equation of the same form,
\begin{equation}
    \frac{\rmd}{\rmd t}\,U_{j+1}(t) = -\rmi H_{j+1}(t)\,U_{j+1}(t)\;,
\end{equation}
with an effective Hamiltonian generated recursively by
\begin{equation}
    H_{j+1}(t) = \rme^{-\ad_{F_j(t)}}H_j(t)
    - \int_0^1\!\rmd x\; \rme^{-x\ad_{F_j(t)}}H_j(t)\;,
    \label{eq:fer-recursion}
\end{equation}
where $\ad_Y(X)\equiv[Y,X]$, conjugation by an exponential has been written
as the exponential of the adjoint action (Appendix~\ref{app:conjugation}),
and the exponents are
\begin{equation}
    F_j(t) = -\rmi\int_0^t\!\rmd s\,H_j(s)\;,
    \label{eq:fer-F}
\end{equation}
with $H_1(t)\equiv H(t)$. This is the defining step of the expansion: at order
$j$ the integrated part of the dynamics is peeled off into $\rme^{F_j(t)}$,
leaving a smaller remainder $H_{j+1}(t)$ for the next step~\cite{fer}. Each
step squares the size of that remainder, which makes the expansion
superconvergent, in the sense of the accelerated perturbation schemes of
Refs.~\cite{Scherer1995,Scherer1998}; its convergence domain is analyzed in
Ref.~\cite{blanes1998}, which for a Hermitian $H(t)$ guarantees convergence
while $\int_0^t\rmd s\,\|H(s)\|<2$, with $\|\cdot\|$ the operator norm; this
radius was recently raised to about $2.6058$~\cite{bagnasacco2026fer}.
Expanding the two exponentials of $\ad_{F_j(t)}$ in Eq.~\eqref{eq:fer-recursion}
and doing the $x$ integral collapses it into a single series of nested
commutators,
\begin{equation}
    H_{j+1}(t) = \sum_{n=0}^{\infty}\frac{(-1)^{n+1}}{(n+2)\,n!}\,
    \ad^{n+1}_{F_j(t)}(H_j(t))\;,
    \label{eq:fer-series}
\end{equation}
with no auxiliary integrals left to perform.

\section{The Fer map on a Lie algebra}
\label{sec:framework}

Two features of the recursion hold for any algebra. We isolate them here,
before specializing to $\mathfrak{su}(2)$.

\subsection{The recursion stays in the algebra}
\label{sec:framework-closure}

Let the initial Hamiltonian be a real combination of Hermitian generators
$\{\mathrm T_a\}_{a=1}^{\dim\mathfrak{g}}$ of a Lie algebra $\mathfrak{g}$,
$[\mathrm{T}_a,\mathrm{T}_b]=\rmi\sum_c \kappa_{abc}\,\mathrm{T}_c$,
\begin{equation}
    H_1(t) = \mathbf{h}_1(t)\cdot\mathbf{T}\;, \qquad
    \mathbf{h}_1(t)\in\mathbb{R}^{\dim\mathfrak{g}}\;,
\end{equation}
and $F_1(t)=-\rmi\,\mathbf{f}_1(t)\cdot\mathbf{T}$, with
$\mathbf{f}_1(t)=\int_0^t\rmd s\,\mathbf{h}_1(s)$. Every term of
Eq.~\eqref{eq:fer-series} is an iterated commutator of $F_1(t)$ with
$H_1(t)$, so the next-order Hamiltonian is again a combination of the same
generators, and by induction
\begin{equation}
    H_j(t) = \mathbf{h}_j(t)\cdot\mathbf{T}\;, \qquad
    F_j(t) = -\rmi\,\mathbf{f}_j(t)\cdot\mathbf{T}\;,
    \label{eq:algebra-closure}
\end{equation}
for all $j\geq 1$. The recursion never leaves $\mathfrak{g}$; it only
reshuffles the real coefficient vector. It therefore closes on the
coefficient vectors themselves, taken as whole functions of time,
\begin{equation}
    \mathbf{h}_j(t) \;\longmapsto\;
    \mathbf{h}_{j+1}(t) = \mathcal{F}[\mathbf{h}_j(t)]\;,
    \label{eq:fer-map}
\end{equation}
with the Fer map $\mathcal{F}$ fixed by the structure constants
$\kappa_{abc}$ of $\mathfrak{g}$ through Eq.~\eqref{eq:fer-series}.
Iterating it generates the chain
\begin{equation}
    \mathbf{h}_1(t) \longmapsto \mathbf{h}_2(t) \longmapsto \cdots
    \longmapsto \mathbf{h}_j(t)
    \label{eq:chain}
\end{equation}
to any desired order. The Fer map is not invertible in
general: if $H_j(t)$ commutes with itself at all times, it also commutes
with its own integral $F_j(t)$, every nested commutator in
Eq.~\eqref{eq:fer-series} vanishes, and $H_{j+1}(t)=0$, so that $\mathcal{F}$
sends all the drives of this kind to the zero drive.

\subsection{Where the chain stops}
\label{sec:framework-solvable}

Whatever the algebra, the evolution generated by
$H(t)=\mathbf{h}(t)\cdot\mathbf{T}$ is solvable in closed form whenever the
Hamiltonian commutes with itself at all times,
\begin{equation}
    [H(t),H(t')] = 0 \qquad \text{for all } t\neq t'\;.
    \label{eq:family-A-def}
\end{equation}
We call family $\mathcal{A}$ the class of drives obeying
Eq.~\eqref{eq:family-A-def}, and we say that the coefficient vector
$\mathbf{h}(t)$ belongs to $\mathcal{A}$ when its Hamiltonian does.

Every Fer chain that reaches family $\mathcal{A}$ stops there, as
anticipated in Sec.~\ref{sec:framework-closure}. Indeed, if
$\mathbf{h}_j(t)\in\mathcal{A}$, the Hamiltonian $H_j(t)$ commutes with its
own integral, Eq.~\eqref{eq:fer-F},
\begin{equation}
    [H_j(t),F_j(t)] = -\rmi\int_0^t\!\rmd s\,[H_j(t),H_j(s)] = 0\;,
    \label{eq:HF-commute}
\end{equation}
because the integrand vanishes identically by
Eq.~\eqref{eq:family-A-def}. The innermost commutator of every term of
Eq.~\eqref{eq:fer-series} is $\ad_{F_j(t)}(H_j(t))=-[H_j(t),F_j(t)]$, so the
whole series vanishes term by term and
\begin{equation}
    \mathcal{F}(\mathcal{A}) = \mathcal{O}\;,
    \qquad \mathcal{O}\equiv\{\bm{0}\}\;.
    \label{eq:A-to-O-general}
\end{equation}
Reaching $\mathcal{A}$ at order $j$ means that the factorization closes
there: $H_{j+1}(t)=0$ and $U(t)=\rme^{F_1(t)}\cdots\rme^{F_j(t)}$ is exact, with
no remainder left. Since $\bm{0}\in\mathcal{A}$, one has
$\mathcal{O}\subseteq\mathcal{A}$, and the chain stays at $\mathcal{O}$
forever after.

The question of Sec.~\ref{sec:intro} is now precise: for which
$\mathbf{h}_1(t)\notin\mathcal{A}$ does the map of Eq.~\eqref{eq:fer-map} reach
$\mathcal{A}$ after finitely many steps? We exclude
$\mathbf{h}_1(t)\in\mathcal{A}$ because it is the trivial case: there
$[H_1(t),F_1(t)]=0$ already at the first step, so $H_j(t)=0$ for all
$j>1$ and the expansion stops after a single factor,
$U(t)=\rme^{F_1(t)}$, which is the time-ordered exponential of
Sec.~\ref{sec:intro} with the time ordering dropped.
Answering the question means going back along the chain, one step at a
time: finding the drives that one step sends into $\mathcal{A}$.

\section{The Fer map on \texorpdfstring{$\mathfrak{su}(2)$}{su(2)}}
\label{sec:su2}

For $\mathfrak{su}(2)$ the map can be written down in closed form. A
component of the Hamiltonian along the identity commutes with
everything: it drops out of every commutator in Eq.~\eqref{eq:fer-series}
and only contributes a global phase to $\rme^{F_1(t)}$, so we take $H(t)$
traceless. We take $\mathbf{T}$ to be a vector of Hermitian matrices
of any dimension, obeying~\cite{sakurai}
\begin{equation}
    [\mathrm{T}_a,\mathrm{T}_b] = 2\rmi\,\varepsilon_{abc}\,\mathrm{T}_c\;,
    \label{eq:su2-commutator}
\end{equation}
that is, twice the spin operators of any nontrivial representation of
$\mathfrak{su}(2)$, for instance the Pauli matrices for a qubit, and write, as
in
Eq.~\eqref{eq:algebra-closure},
\begin{equation}
    H_j(t) = \mathbf{h}_j(t)\cdot\mathbf{T}\;, \qquad
    F_j(t) = -\rmi\,\mathbf{f}_j(t)\cdot\mathbf{T}\;,
    \label{eq:su2-decomposition}
\end{equation}
with $\mathbf{h}_j(t)$ a real three-dimensional vector, the drive vector
about which the mean spin precesses, as the Bloch vector of a qubit in
the representation of Feynman,
Vernon, and Hellwarth~\cite{feynman}, and
$\mathbf{f}_j(t)=\int_0^t\rmd s\,\mathbf{h}_j(s)$. The norms of the two
vectors are abbreviated as
$\mathrm{f}_j(t)\equiv\|\mathbf{f}_j(t)\|$ and
$\mathrm{h}_j(t)\equiv\|\mathbf{h}_j(t)\|$, with $\|\cdot\|$ the Euclidean
norm. By construction
\begin{equation}
    \dot{\mathbf{f}}_j(t) = \mathbf{h}_j(t)\;, \qquad \mathbf{f}_j(0)=\bm{0}\;.
    \label{eq:f-dot-equals-h}
\end{equation}

Since the commutator is bilinear, Eq.~\eqref{eq:su2-commutator} turns
the commutator of two combinations of the $\mathrm{T}_a$ into the cross
product of their coefficient vectors,
\begin{equation}
    [\mathbf{f}_j(t)\cdot\mathbf{T},\,\mathbf{h}_j(t)\cdot\mathbf{T}]
    = 2\rmi\,\Lambda_{\mathbf{f}_j(t)}(\mathbf{h}_j(t))\cdot\mathbf{T}\;,
    \label{eq:su2-commutator-vector}
\end{equation}
where $\Lambda_{\mathbf{x}}(\mathbf{y})\equiv\mathbf{x}\times\mathbf{y}$, so
that the operator series of Eq.~\eqref{eq:fer-series} becomes a recursion
entirely
in $\mathbb{R}^3$,
\begin{equation}
    \mathbf{h}_{j+1}(t) = \sum_{n=0}^{\infty}\Phi_n\,
    \Lambda^{n+1}_{\mathbf{f}_j(t)}\bigl(\mathbf{h}_j(t)\bigr)\;,
    \label{eq:su2-fer-map-series}
\end{equation}
with numerical coefficients $\Phi_n\equiv(-2)^{n+1}/[(n+2)\,n!]$.

This series resums into elementary functions: summing its odd and even
powers separately gives the closed map
\begin{equation}
    \mathbf{h}_{j+1}(t) = \alpha(\mathrm{f}_j(t))\,\mathbf{p}_{j+1}(t)
    + \beta(\mathrm{f}_j(t))\,\mathbf{q}_{j+1}(t)\;,
    \label{eq:su2-fer-map-closed}
\end{equation}
built from the two auxiliary vectors
\begin{align}
    \mathbf{p}_{j+1}(t) &\equiv \mathbf{f}_j(t)\times\mathbf{h}_j(t)\;, \notag\\
    \mathbf{q}_{j+1}(t) &\equiv \mathbf{p}_{j+1}(t)\times\mathbf{f}_j(t)\;,
    \label{eq:su2-aux-vectors}
\end{align}
and weighted by the two scalar functions
\begin{align}
    \alpha(x) &= \frac{1-\cos(2x)}{2x^2} - \frac{\sin(2x)}{x}\;,
    \label{eq:alpha-def}\\
    \beta(x) &= \frac{\cos(2x)}{x^2} - \frac{\sin(2x)}{2x^3}\;.
    \label{eq:beta-def}
\end{align}
Equation~\eqref{eq:su2-fer-map-closed} is the closed-form realization of the
abstract map of Eq.~\eqref{eq:fer-map} for $\mathfrak{su}(2)$: it is what we
iterate
from now on. Appendix~\ref{app:closed-form} derives it twice, once by
resumming the series of Eq.~\eqref{eq:su2-fer-map-series} and once,
independently,
by inserting the Rodrigues rotation formula~\cite{sakurai} into the
recursion of Eq.~\eqref{eq:fer-recursion} and integrating.

Each factor of the product of Eq.~\eqref{eq:fer-product} is an explicit
rotation: $\rme^{F_j(t)}=\rme^{-\rmi\mathbf{f}_j(t)\cdot\mathbf{T}}$ rotates the
spin by the angle $2\mathrm{f}_j(t)$ about the unit vector
$\hat{\mathbf{f}}_j(t)\equiv\mathbf{f}_j(t)/\mathrm{f}_j(t)$, and in any
representation it is given by the corresponding rotation
matrix~\cite{sakurai}. For a qubit, with $\mathbf{T}=\bm{\sigma}$ the vector
of Pauli matrices,
$\rme^{F_j(t)}=\cos\mathrm{f}_j(t)\,\openone
-\rmi\sin\mathrm{f}_j(t)\,\hat{\mathbf{f}}_j(t)\cdot\bm{\sigma}$.

Two properties of the map will be used repeatedly. First, both auxiliary
vectors lie in the plane $\Pi_j(t)$ orthogonal to $\mathbf{f}_j(t)$,
Fig.~\ref{fig:fer-step}: $\mathbf{p}_{j+1}(t)$ is orthogonal to both
$\mathbf{f}_j(t)$ and $\mathbf{h}_j(t)$, and $\mathbf{q}_{j+1}(t)$ is
$\mathrm{f}_j^2(t)$ times the projection of $\mathbf{h}_j(t)$ onto
$\Pi_j(t)$. Hence the whole output lies in $\Pi_j(t)$,
\begin{equation}
    \mathbf{h}_{j+1}(t)\perp\mathbf{f}_j(t)\;.
    \label{eq:su2-orthogonality}
\end{equation}
Second, $\mathbf{f}_j(0)=\bm{0}$ makes both auxiliary vectors vanish at the
origin, so every Fer output starts there,
\begin{equation}
    \mathbf{h}_{j+1}(0)=\bm{0}\;,
    \label{eq:su2-vanishing-at-origin}
\end{equation}
a constraint that the initial Hamiltonian $\mathbf{h}_1(t)$ does not carry.

\section{Rigidity of the \texorpdfstring{$\mathfrak{su}(2)$}{su(2)} chain}
\label{sec:su2-rigidity}

With the map available in closed form, the question of
Sec.~\ref{sec:framework-solvable} can be answered by inspection of a single
backward step. We do it first for family $\mathcal{A}$, the drives that
commute with themselves at all times, and then for the uniformly precessing
drives of Rabi's rotating field, a textbook solvable class outside
$\mathcal{A}$. In both cases the answer is negative: the recursion reaches the
class at order $j>1$ only when the original drive was already in
$\mathcal{A}$, and then only through the zero drive.

Throughout this section the original drive $\mathbf{h}_1(t)$ is taken to be
real-analytic in $t$. The weights $\alpha$ and $\beta$ of
Eqs.~\eqref{eq:alpha-def} and~\eqref{eq:beta-def} are even, and their
singularities at the origin are removable, hence they are analytic functions
of $\mathbf{f}_j(t)\cdot\mathbf{f}_j(t)$, so every
$\mathbf{h}_j(t)$ produced by Eq.~\eqref{eq:su2-fer-map-closed} is
real-analytic as well.

\subsection{Drives with a fixed direction}

Read through the commutator of Eq.~\eqref{eq:su2-commutator-vector}, the
defining condition of family $\mathcal{A}$, Eq.~\eqref{eq:family-A-def}, becomes
\begin{equation}
    \mathbf{h}(t) \times \mathbf{h}(t') = \bm{0}
    \qquad \text{for all } t\neq t'\;,
    \label{eq:su2-family-A}
\end{equation}
i.e., the trajectory of $\mathbf{h}(t)$ never leaves a fixed direction in
$\mathbb{R}^3$: it moves along a single line through the origin.

Since $\mathcal{F}(\mathcal{A})=\mathcal{O}\subseteq\mathcal{A}$ by
Eq.~\eqref{eq:A-to-O-general}, the following theorem amounts to a single
step back along the chain: a drive that one step sends into $\mathcal{A}$ was
already in $\mathcal{A}$.

\begin{theorem}
\label{thm:su2-rigidity}
Let the original drive $\mathbf{h}_1(t)$ be real-analytic. In the Fer
expansion associated with the $\mathfrak{su}(2)$ algebra, if the coefficient
vector $\mathbf{h}_j(t)$ at some order $j>1$ belongs to family~$\mathcal{A}$,
then $\mathbf{h}_j(t)\equiv\bm{0}$, and the coefficient vector
$\mathbf{h}_1(t)$ of the original Hamiltonian
$H_1(t)=\mathbf{h}_1(t)\cdot\mathbf{T}$ belongs to family~$\mathcal{A}$ as
well.
\end{theorem}

\begin{proof}[Proof of Theorem~\ref{thm:su2-rigidity}]
We prove the one-step implication, valid for every $j\geq1$, from
$\mathbf{h}_{j+1}(t)\in\mathcal{A}$
to $\mathbf{h}_{j+1}(t)\equiv\bm{0}$ and $\mathbf{h}_j(t)\in\mathcal{A}$, and
conclude by descending induction along the chain of Eq.~\eqref{eq:chain}.

Assume $\mathbf{h}_{j+1}(t)\in\mathcal{A}$: both $\mathbf{h}_{j+1}(t)$ and
$\dot{\mathbf{h}}_{j+1}(t)$ lie on a fixed line through the origin. By
Eq.~\eqref{eq:su2-orthogonality}, $\mathbf{f}_j(t)$ is orthogonal to this
line wherever $\mathbf{h}_{j+1}(t)\neq\bm{0}$, hence at all times unless
$\mathbf{h}_{j+1}(t)\equiv\bm{0}$, since the zeros of a real-analytic
function are isolated. Either way
$\dot{\mathbf{h}}_{j+1}(t)\perp\mathbf{f}_j(t)$. Differentiating
Eq.~\eqref{eq:su2-fer-map-closed} and projecting on $\mathbf{f}_j(t)$
(Appendix~\ref{app:derivative}), every term is a cross product with
$\mathbf{f}_j(t)$ and drops out, except one of the two generated by
$\dot{\mathbf{q}}_{j+1}(t)$. What survives is the necessary condition
\begin{equation}
    \beta\bigl(\mathrm{f}_j(t)\bigr)\,\mathbf{f}_j(t)\cdot
    \bigl[\mathbf{p}_{j+1}(t)\times\mathbf{h}_j(t)\bigr] = 0\;.
    \label{eq:su2-necessary-closed}
\end{equation}
By the triple product identity and the Lagrange identity,
\begin{align}
    \mathbf{f}_j(t)\cdot\bigl[\mathbf{p}_{j+1}(t)\times\mathbf{h}_j(t)\bigr]
    &= \bigl(\mathbf{f}_j(t)\cdot\mathbf{h}_j(t)\bigr)^2
       - \mathrm{f}_j^2(t)\,\mathrm{h}_j^2(t) \notag\\
    &= -\bigl\|\mathbf{f}_j(t)\times\mathbf{h}_j(t)\bigr\|^2\;,
\end{align}
so that Eq.~\eqref{eq:su2-necessary-closed} reads
\begin{equation}
    \beta\bigl(\mathrm{f}_j(t)\bigr)\,
    \bigl\|\mathbf{f}_j(t)\times\mathbf{h}_j(t)\bigr\|^2 = 0\;.
    \label{eq:su2-necessary-final}
\end{equation}

At every instant this holds either because $\beta(\mathrm{f}_j(t))=0$ or
because $\mathbf{f}_j(t)\times\mathbf{h}_j(t)=\bm{0}$, and the first
alternative can only occur at isolated times. Indeed, as noted at the
beginning of this section, the composition $t\mapsto\beta(\mathrm{f}_j(t))$ is
real-analytic, even though the norm $\mathrm{f}_j(t)$ itself need not be where
$\mathbf{f}_j(t)$ vanishes. Near the origin,
\begin{equation}
    \beta(x) = -\tfrac{4}{3} + \tfrac{8}{15}x^2 + O(x^4)\;,
    \qquad \beta(0) = -\tfrac{4}{3}\neq0\;,
    \label{eq:beta-at-origin}
\end{equation}
and since $\mathbf{f}_j(0)=\bm{0}$, the composition equals
$\beta(0)=-4/3$ at $t=0$, so it is not identically zero, and its zeros are
isolated. Equation~\eqref{eq:su2-necessary-final} then forces
$\mathbf{f}_j(t)\times\mathbf{h}_j(t)=\bm{0}$ on the complement of a discrete
set, which is dense, and by continuity everywhere:
\begin{equation}
    \mathbf{h}_j(t) \parallel \mathbf{f}_j(t)
    \qquad \text{for all } t\;,
    \label{eq:su2-f-parallel-h}
\end{equation}
that is, $\mathbf{f}_j(t)\times\mathbf{h}_j(t)\equiv\bm{0}$.

Two consequences follow. First, both auxiliary vectors of
Eq.~\eqref{eq:su2-aux-vectors} vanish identically, and with them the output
of Eq.~\eqref{eq:su2-fer-map-closed}: $\mathbf{h}_{j+1}(t)\equiv\bm{0}$.
Second, since $\dot{\mathbf{f}}_j(t)=\mathbf{h}_j(t)$,
Eq.~\eqref{eq:su2-f-parallel-h} reads
$\mathbf{f}_j(t)\times\dot{\mathbf{f}}_j(t)\equiv\bm{0}$. If
$\mathbf{f}_j(t)\equiv\bm{0}$, then $\mathbf{h}_j(t)\equiv\bm{0}$. Otherwise,
on any interval where $\mathbf{f}_j(t)\neq\bm{0}$, its derivative is
parallel to it and its direction does not change; being real-analytic,
$\mathbf{f}_j(t)$ then stays on that line at all times, and so does
$\mathbf{h}_j(t)=\dot{\mathbf{f}}_j(t)$. This is
$\mathbf{h}_j(t)\in\mathcal{A}$.
\end{proof}

\begin{remark}
The proof never inverts $\mathcal{F}$, which in general cannot be done
(Sec.~\ref{sec:framework-closure}). It does not reconstruct $\mathbf{h}_j(t)$
from $\mathbf{h}_{j+1}(t)$: it takes any drive $\mathbf{h}_j(t)$ that one step
sends into $\mathcal{A}$ and constrains it using only the forward map,
Eq.~\eqref{eq:su2-fer-map-closed}.
\end{remark}

Together with Eq.~\eqref{eq:A-to-O-general},
Theorem~\ref{thm:su2-rigidity} says that the drives that one step sends
into $\mathcal{A}$, or into $\mathcal{O}$, are exactly those already in
$\mathcal{A}$; in particular, no real-analytic drive outside
$\mathcal{A}$ produces, at any later order, an effective Hamiltonian that
commutes with itself at all times.

\subsection{Precessing drives}
\label{sec:su2-family-precessing}

Family $\mathcal{A}$ is not the only solvable class in $\mathfrak{su}(2)$.
After $j$ steps the propagator reads, by Eq.~\eqref{eq:fer-product},
$U(t)=\rme^{F_1(t)}\cdots\rme^{F_j(t)}\,U_{j+1}(t)$, with $U_{j+1}(t)$
generated by $H_{j+1}(t)$. The product terminates when $H_{j+1}(t)$ vanishes,
which involves only family $\mathcal{A}$, Eq.~\eqref{eq:A-to-O-general}, and is
settled by Theorem~\ref{thm:su2-rigidity}. The expansion can be stopped just
as well whenever $H_{j+1}(t)$ belongs to a class whose dynamics is solved
exactly: $U_{j+1}(t)$, and with it the whole propagator, is then known in
closed form. Here we focus on the best-known such class, Rabi's rotating
field~\cite{rabi}, which we call family $\widetilde{\mathcal{A}}$: it collects
the drives whose vector precesses about a fixed unit axis $\hat{\mathbf{n}}$
with constant angular velocity $\omega$,
\begin{align}
    \mathbf{h}(t) = {}&\cos(\omega t)\,\mathbf{h}(0)
    + \sin(\omega t)\,\bigl(\mathbf{h}(0)\times\hat{\mathbf{n}}\bigr)
    \notag\\
    &{}+ \bigl[1-\cos(\omega t)\bigr]\,
    \bigl(\hat{\mathbf{n}}\cdot\mathbf{h}(0)\bigr)\,\hat{\mathbf{n}}\;.
    \label{eq:family-A-precessing-def}
\end{align}
Despite the richer time dependence, these drives are solvable too: in the frame
rotating about
$\hat{\mathbf{n}}$, the propagator generated by
$H(t)=\mathbf{h}(t)\cdot\mathbf{T}$ factorizes as
\begin{equation}
    U(t) = \rme^{\rmi(\omega t/2)\hat{\mathbf{n}}\cdot\mathbf{T}}\,
    \widetilde{U}(t)\;,
\end{equation}
where $\widetilde{U}$ is generated by the time-independent Hamiltonian
$\widetilde{H}=\bigl(\mathbf{h}(0)+(\omega/2)\,\hat{\mathbf{n}}\bigr)
\cdot\mathbf{T}$. Yet the chain cannot reach $\widetilde{\mathcal{A}}$ either.

\begin{theorem}
\label{thm:su2-precessing}
Let the original drive $\mathbf{h}_1(t)$ be real-analytic. In the Fer
expansion associated with the $\mathfrak{su}(2)$ algebra, if the coefficient
vector $\mathbf{h}_j(t)$ at some order $j>1$ belongs to family
$\widetilde{\mathcal{A}}$, then $\mathbf{h}_j(t)\equiv\bm{0}$, and the
coefficient vector $\mathbf{h}_1(t)$ of the original Hamiltonian
$H_1(t)=\mathbf{h}_1(t)\cdot\mathbf{T}$ belongs to family~$\mathcal{A}$.
\end{theorem}

\begin{proof}[Proof of Theorem~\ref{thm:su2-precessing}]
Since $j>1$, $\mathbf{h}_j(t)$ is a genuine Fer output and therefore
vanishes at $t=0$ by Eq.~\eqref{eq:su2-vanishing-at-origin}. A
member of $\widetilde{\mathcal{A}}$ has constant magnitude
$\|\mathbf{h}(t)\|=\|\mathbf{h}(0)\|$ by
Eq.~\eqref{eq:family-A-precessing-def}, so one that vanishes at the origin
vanishes identically. Hence $\mathbf{h}_j(t)\equiv\bm{0}\in\mathcal{A}$, and
Theorem~\ref{thm:su2-rigidity} gives $\mathbf{h}_1(t)\in\mathcal{A}$.
\end{proof}

\section{Conclusion}
\label{sec:conclusion}

For a spin of any size, and in every representation of
$\mathfrak{su}(2)$, we have summed the series of nested commutators that
defines each step of the Fer expansion into a closed map on time-dependent
vectors in $\mathbb{R}^3$, with two elementary trigonometric weights. Each factor of the
product follows from the previous one through a single time integral and
elementary functions of the norm of the previous integral,
$\mathrm{f}_j(t)=\|\mathbf{f}_j(t)\|$. The known strengths of the expansion
thus become directly usable. The product is unitary at every truncation, and
each factor squares the size of the remainder, so that the error decreases
doubly exponentially with the number of factors~\cite{blanes2009}, within a
convergence domain that has recently been enlarged by about
$30\%$~\cite{bagnasacco2026fer}. A few factors therefore give an accurate
propagator for any spin drive, such as the pulse sequences of magnetic
resonance~\cite{haeberlen1968} and dynamical decoupling~\cite{viola1999}, or
the control pulses that implement quantum gates~\cite{koch2022}. With the
closed form obtained here and the recently enlarged convergence domain, the
Fer product is thus well suited also to quantum algorithms, in which product
formulas are among the most widely used methods to simulate time
evolution~\cite{lloyd1996,childs2021}: a truncated Fer product is a product
formula of explicit rotations, which can be applied directly as gates.

We have also used the closed map to ask whether the Fer product can be
stopped after finitely many factors on one of two classes of effective
Hamiltonians whose evolution is known in closed form: those that commute
with themselves at all times, on which the product terminates, and those
that precess about a fixed axis with constant angular velocity, as in Rabi's
rotating field. For a real-analytic drive, no effective Hamiltonian beyond
the first falls into either class unless the first exponential is already
exact.
In $\mathfrak{su}(2)$, exactly solvable drives therefore still have to be
engineered backward, from the solution to the
drive~\cite{Barnes2012,Messina2014,GueryOdelin2019}, and whether the chain
can reach other solvable classes remains open. Running the Fer map until the
effective Hamiltonian becomes exactly solvable is nonetheless a direct method
to search for solvable dynamics, since it starts from the drive rather than
from the solution. For the families considered in this work, it fails for $\mathfrak{su}(2)$, but the first part of our
analysis, in Sec.~\ref{sec:framework}, holds for any finite-dimensional Lie
algebra, so the method can be applied to other algebras, for instance those
that describe squeezing and parametric amplification in quantum
optics~\cite{qvarfort}, where the answer may differ.

\begin{acknowledgments}
The authors thank Riccardo Castellano (Department of Applied Physics,
University of Geneva) for helpful discussions on the presentation of the
introduction and conclusions.
\end{acknowledgments}

\section*{Use of AI}
The authors used Claude Code (Anthropic) to assist
with spelling corrections, graphical refinement of figure~\ref{fig:fer-step},
and searches for relevant literature. The scientific
content, including the theorems and their proofs, is the authors’ own. The authors reviewed the output and take full responsibility for the manuscript.

\section*{Data availability}
No data were created or analyzed in this article.

\onecolumngrid
\appendix

\section{Conjugation formula}
\label{app:conjugation}

Consider two operators $X,Y$ and the one-parameter family
$g(\lambda)\equiv\rme^{\lambda Y}X\rme^{-\lambda Y}$ with
$\lambda\in\mathbb{R}$. Differentiating $n$ times gives
\begin{equation}
    \frac{\rmd^n}{\rmd \lambda^n}\,g(\lambda)
    = \rme^{\lambda Y}\,\ad_Y^n(X)\,\rme^{-\lambda Y}\;,
    \qquad \ad_Y(X)\equiv[Y,X]\;,
\end{equation}
as follows by induction, since one further derivative produces one further
commutator with $Y$ between the same two exponentials. Expanding
$g(\lambda)$ in a Taylor series around $\lambda=0$ and evaluating it at
$\lambda=\gamma$ then yields
\begin{equation}\label{eq:conjugation-identity-operator}
    \rme^{\gamma Y}\,X\,\rme^{-\gamma Y}
    = \sum_{n=0}^{\infty}\frac{\gamma^n}{n!}\,\ad_Y^n(X)\;,
\end{equation}
valid for arbitrary $X,Y$. This is what allows conjugation by an exponential
to be written as the exponential of the adjoint action, in the
recursion of Eq.~\eqref{eq:fer-recursion} as in its double-integral
form, Eq.~\eqref{eq:fer-recursion-double}.

We now translate Eq.~\eqref{eq:conjugation-identity-operator} into vector
form, with $X=\mathbf{x}\cdot\mathbf{T}$ and $Y=\rmi\,\mathbf{y}\cdot\mathbf{T}$
for real $\mathbf{x},\mathbf{y}$ and Hermitian generators $\mathbf{T}$. By the
$\mathfrak{su}(2)$ commutator of Eq.~\eqref{eq:su2-commutator}, a single
application
of $\ad_Y$ gives
\begin{equation}
    \ad_Y(X) = \rmi\,[\mathbf{y}\cdot\mathbf{T},\mathbf{x}\cdot\mathbf{T}]
    = \rmi\cdot 2\rmi\,\Lambda_{\mathbf{y}}(\mathbf{x})\cdot\mathbf{T}
    = -2\,\Lambda_{\mathbf{y}}(\mathbf{x})\cdot\mathbf{T}\;.
\end{equation}
The result is again of the form $\mathbf{z}\cdot\mathbf{T}$, so the same
computation iterates: by linearity of $\Lambda_{\mathbf{y}}$, each further
application of $\ad_Y$ contributes one factor of $-2$ and one more power of
$\Lambda_{\mathbf{y}}$, whence
$\ad_Y^n(X)=(-2)^n\Lambda_{\mathbf{y}}^n(\mathbf{x})\cdot\mathbf{T}$
by induction. Substituting into
Eq.~\eqref{eq:conjugation-identity-operator} gives
\begin{equation}
    \rme^{\rmi\gamma\mathbf{y}\cdot\mathbf{T}}\,(\mathbf{x}\cdot\mathbf{T})\,
    \rme^{-\rmi\gamma\mathbf{y}\cdot\mathbf{T}}
    = \sum_{n=0}^{\infty}\frac{(-2\gamma)^n}{n!}\,
    \Lambda_{\mathbf{y}}^n(\mathbf{x})\cdot\mathbf{T}\;,
    \qquad
    \Lambda_{\mathbf{y}}(\mathbf{x}) \equiv \mathbf{y}\times\mathbf{x}\;.
    \label{eq:conjugation-rescaled}
\end{equation}

\section{Closed form of the Fer map on
\texorpdfstring{$\mathfrak{su}(2)$}{su(2)}}
\label{app:closed-form}

Here we derive Eq.~\eqref{eq:su2-fer-map-closed} in two independent ways.

\subsection{From the series}

The key structural fact is that $\Lambda_{\mathbf{f}_j(t)}$ generates only two
directions out of $\mathbf{h}_j(t)$: every power
$\Lambda^n_{\mathbf{f}_j(t)}(\mathbf{h}_j(t))$ with $n\geq1$ is a multiple of
either $\mathbf{p}_{j+1}(t)=\mathbf{f}_j(t)\times\mathbf{h}_j(t)$ or
$\mathbf{q}_{j+1}(t)=
(\mathbf{f}_j(t)\times\mathbf{h}_j(t))\times\mathbf{f}_j(t)$.
Indeed, from
$\Lambda_{\mathbf{f}_j(t)}(\mathbf{h}_j(t))=
\mathbf{f}_j(t)\times\mathbf{h}_j(t)$, a
second application and the BAC--CAB identity give
\begin{equation}
    \Lambda_{\mathbf{f}_j(t)}^2(\mathbf{h}_j(t))
    = \mathbf{f}_j(t)\times\bigl(\mathbf{f}_j(t)\times\mathbf{h}_j(t)\bigr)
    = \mathbf{f}_j(t)\bigl(\mathbf{f}_j(t)\cdot\mathbf{h}_j(t)\bigr)
    - \mathrm{f}_j^2(t)\,\mathbf{h}_j(t)\;,
\end{equation}
and a third, using $\mathbf{f}_j(t)\times\mathbf{f}_j(t)=\bm{0}$,
\begin{equation}
    \Lambda_{\mathbf{f}_j(t)}^3(\mathbf{h}_j(t))
    = \bigl(\mathbf{f}_j(t)\cdot\mathbf{h}_j(t)\bigr)
    \bigl(\mathbf{f}_j(t)\times\mathbf{f}_j(t)\bigr)
    - \mathrm{f}_j^2(t)\bigl(\mathbf{f}_j(t)\times\mathbf{h}_j(t)\bigr)
    = -\mathrm{f}_j^2(t)\bigl(\mathbf{f}_j(t)\times\mathbf{h}_j(t)\bigr)\;.
\end{equation}
Each further application swaps the two structures, and every second one
brings a factor $-\mathrm{f}_j^2(t)$, so the pattern closes at all orders,
\begin{align}
    \Lambda_{\mathbf{f}_j(t)}^{2m+1}(\mathbf{h}_j(t))
    &= \bigl[-\mathrm{f}_j^2(t)\bigr]^m\,\mathbf{p}_{j+1}(t)\;,
    \label{eq:su2-lambda-odd}\\[2pt]
    \Lambda_{\mathbf{f}_j(t)}^{2m+2}(\mathbf{h}_j(t))
    &= \bigl[-\mathrm{f}_j^2(t)\bigr]^{m}\,
    \mathbf{f}_j(t)\times\mathbf{p}_{j+1}(t)
    = -\bigl[-\mathrm{f}_j^2(t)\bigr]^{m}\,\mathbf{q}_{j+1}(t)\;,
    \label{eq:su2-lambda-even}
\end{align}
with $\mathbf{p}_{j+1}(t)$ and $\mathbf{q}_{j+1}(t)$ the auxiliary vectors of
Eq.~\eqref{eq:su2-aux-vectors}. Splitting the
series of Eq.~\eqref{eq:su2-fer-map-series} into even and odd parts and
inserting
Eqs.~\eqref{eq:su2-lambda-odd} and~\eqref{eq:su2-lambda-even},
\begin{equation}
    \mathbf{h}_{j+1}(t)
    = \left(\sum_{m=0}^\infty \Phi_{2m}\bigl[-\mathrm{f}_j^2(t)\bigr]^m\right)
    \mathbf{p}_{j+1}(t)
    - \left(\sum_{m=0}^\infty
    \Phi_{2m+1}\bigl[-\mathrm{f}_j^2(t)\bigr]^{m}\right)
    \mathbf{q}_{j+1}(t)\;,
\end{equation}
and the two sums are precisely the Taylor series of the elementary functions
of Eqs.~\eqref{eq:alpha-def} and~\eqref{eq:beta-def},
\begin{equation}
    \sum_{m=0}^\infty \Phi_{2m}\bigl[-\mathrm{f}_j^2(t)\bigr]^m =
    \alpha(\mathrm{f}_j(t))\;,
    \qquad
    -\sum_{m=0}^\infty \Phi_{2m+1}\bigl[-\mathrm{f}_j^2(t)\bigr]^m =
    \beta(\mathrm{f}_j(t))\;,
\end{equation}
which is Eq.~\eqref{eq:su2-fer-map-closed}. In particular $\alpha$ and
$\beta$ are the closed-form sums of the even and odd subseries of the
recursion, and the apparent singularities at $\mathrm{f}_j(t)=0$ are removable.

\subsection{From the Rodrigues formula}

The same result follows from the recursion of Eq.~\eqref{eq:fer-recursion}
without
expanding in commutators. Written as a double integral, equivalent to
Eq.~\eqref{eq:fer-recursion} since both reduce to the
series of Eq.~\eqref{eq:fer-series}, the recursion reads
\begin{equation}
    H_{j+1}(t) = \int_0^1\!\rmd x\int_0^{x}\!\rmd y\;
    \rme^{-(1-y)F_j(t)}\,[H_j(t),F_j(t)]\,\rme^{(1-y)F_j(t)}\;.
    \label{eq:fer-recursion-double}
\end{equation}
The conjugation of Eq.~\eqref{eq:conjugation-rescaled} resums, for
$\mathfrak{su}(2)$, into the Rodrigues rotation formula~\cite{sakurai}
\begin{equation}
    \rme^{\rmi\mathbf{y}\cdot\mathbf{T}}(\mathbf{x}\cdot\mathbf{T})
    \rme^{-\rmi\mathbf{y}\cdot\mathbf{T}}
    = \Bigl[\cos(2\|\mathbf{y}\|)\,\mathbf{x}
    + \frac{\sin(2\|\mathbf{y}\|)}{\|\mathbf{y}\|}\,(\mathbf{x}\times\mathbf{y})
    + \frac{1-\cos(2\|\mathbf{y}\|)}{\|\mathbf{y}\|^2}\,
    (\mathbf{x}\cdot\mathbf{y})\,\mathbf{y}\Bigr]\cdot\mathbf{T}\;,
    \label{eq:rodrigues}
\end{equation}
where $\|\cdot\|$ is the Euclidean norm in $\mathbb{R}^3$. Substituting
Eq.~\eqref{eq:rodrigues} into Eq.~\eqref{eq:fer-recursion-double}, carrying
out the integrations over the auxiliary variables and using the orthogonality
$(\mathbf{f}_j(t)\times\mathbf{h}_j(t))\cdot\mathbf{f}_j(t)=0$ gives
\begin{equation}
\begin{split}
    H_{j+1}(t) = -2\Biggl[\,
    &\mathbf{p}_{j+1}(t)
    \left(\frac{\sin(2\mathrm{f}_j(t))}{2\mathrm{f}_j(t)}
    + \frac{\cos(2\mathrm{f}_j(t))-1}{4\mathrm{f}_j^2(t)}\right)\\
    &+\frac{\mathbf{q}_{j+1}(t)}{\mathrm{f}_j(t)}
    \left(\frac{\sin(2\mathrm{f}_j(t))}{4\mathrm{f}_j^2(t)}
    -
    \frac{\cos(2\mathrm{f}_j(t))}{2\mathrm{f}_j(t)}\right)\Biggr]
    \cdot\mathbf{T}\;,
\end{split}
    \label{eq:su2-H-next}
\end{equation}
whose vector of coefficients is again Eq.~\eqref{eq:su2-fer-map-closed} with
the weights of Eqs.~\eqref{eq:alpha-def} and~\eqref{eq:beta-def}. This is an
independent check of the resummation above.

\section{Derivative of the Fer output}
\label{app:derivative}

The proof of Theorem~\ref{thm:su2-rigidity} needs the time derivative of the
closed map of Eq.~\eqref{eq:su2-fer-map-closed}. From
$\dot{\mathbf{f}}_j(t)=\mathbf{h}_j(t)$
and $\mathbf{h}_j(t)\times\mathbf{h}_j(t)=\bm{0}$, the auxiliary vectors of
Eq.~\eqref{eq:su2-aux-vectors} obey
\begin{equation}
    \dot{\mathbf{p}}_{j+1}(t) = \mathbf{f}_j(t)\times\dot{\mathbf{h}}_j(t)\;,
    \qquad
    \dot{\mathbf{q}}_{j+1}(t) =
    \bigl(\mathbf{f}_j(t)\times\dot{\mathbf{h}}_j(t)\bigr)\times\mathbf{f}_j(t)
    + \bigl(\mathbf{f}_j(t)\times\mathbf{h}_j(t)\bigr)\times\mathbf{h}_j(t)\;,
\end{equation}
so that differentiating Eq.~\eqref{eq:su2-fer-map-closed} gives
\begin{align}
    \dot{\mathbf{h}}_{j+1}(t)
    ={}& \bigl(\mathbf{f}_j(t)\times\dot{\mathbf{h}}_j(t)\bigr)\,
         \alpha\bigl(\mathrm{f}_j(t)\bigr)
       + \mathbf{p}_{j+1}(t)\,
         \frac{\alpha'\bigl(\mathrm{f}_j(t)\bigr)}{\mathrm{f}_j(t)}\,
         \bigl(\mathbf{f}_j(t)\cdot\mathbf{h}_j(t)\bigr)
    \notag\\
    &+
    \bigl[\bigl(\mathbf{f}_j(t)\times\dot{\mathbf{h}}_j(t)\bigr)
       \times\mathbf{f}_j(t)
       +
       \bigl(\mathbf{f}_j(t)\times\mathbf{h}_j(t)\bigr)
       \times\mathbf{h}_j(t)\bigr]\,
       \beta\bigl(\mathrm{f}_j(t)\bigr)
    \notag\\
    &+ \mathbf{q}_{j+1}(t)\,
       \frac{\beta'\bigl(\mathrm{f}_j(t)\bigr)}{\mathrm{f}_j(t)}\,
       \bigl(\mathbf{f}_j(t)\cdot\mathbf{h}_j(t)\bigr)\;.
    \label{eq:su2-h-next-derivative-closed}
\end{align}
Here a prime denotes the derivative of a function with respect to its
argument, $\alpha'(x)\equiv\rmd\alpha(x)/\rmd x$, and we used
$\rmd\,\alpha(\mathrm{f}_j(t))/\rmd t
=[\alpha'(\mathrm{f}_j(t))/\mathrm{f}_j(t)]\,
\mathbf{f}_j(t)\cdot\dot{\mathbf{f}}_j(t)$, and likewise for $\beta$. Since
$\alpha$ and $\beta$ are even and regular at the origin, so are
$\alpha'(x)/x$ and $\beta'(x)/x$, and
Eq.~\eqref{eq:su2-h-next-derivative-closed}
stays finite also where $\mathbf{f}_j(t)$ vanishes. Every term on the right-hand
side is a cross product with $\mathbf{f}_j(t)$, and
therefore orthogonal to it, with the single exception of
$(\mathbf{f}_j(t)\times\mathbf{h}_j(t))\times\mathbf{h}_j(t)$. Projecting
Eq.~\eqref{eq:su2-h-next-derivative-closed} on $\mathbf{f}_j(t)$ thus leaves
\begin{equation}
    \mathbf{f}_j(t)\cdot\dot{\mathbf{h}}_{j+1}(t)
    = \beta(\mathrm{f}_j(t))\;\mathbf{f}_j(t)\cdot
    \bigl[(\mathbf{f}_j(t)\times\mathbf{h}_j(t))\times\mathbf{h}_j(t)\bigr]\;,
\end{equation}
which vanishes whenever $\mathbf{h}_{j+1}(t)\in\mathcal{A}$, since then
$\dot{\mathbf{h}}_{j+1}(t)\perp\mathbf{f}_j(t)$, as shown in the proof of
Theorem~\ref{thm:su2-rigidity}. This is
Eq.~\eqref{eq:su2-necessary-closed}.

\twocolumngrid
\bibliographystyle{apsrev4-2}
\bibliography{references}

@Article{dyson1949radiation,
  author  = {Dyson, F. J.},
  journal = {Phys. Rev.},
  title   = {The Radiation Theories of {T}omonaga, {S}chwinger, and {F}eynman},
  year    = {1949},
  volume  = {75}, number = {3}, pages = {486--502},
  doi     = {10.1103/PhysRev.75.486},
  url     = {https://doi.org/10.1103/PhysRev.75.486},
}

@Article{dyson,
  author    = {Dyson, F. J.},
  journal   = {Phys. Rev.},
  title     = {The ${S}$ Matrix in Quantum Electrodynamics},
  year      = {1949},
  issn      = {0031-899X},
  month     = jun,
  number    = {11},
  pages     = {1736--1755},
  volume    = {75},
  doi       = {10.1103/physrev.75.1736},
  publisher = {American Physical Society (APS)},
  url       = {https://doi.org/10.1103/PhysRev.75.1736},
}

@Article{rau-unitary,
  author    = {Rau, A. R. P.},
  journal   = {Phys. Rev. Lett.},
  title     = {Unitary Integration of Quantum {L}iouville-{B}loch Equations},
  year      = {1998},
  number    = {22},
  pages     = {4785--4789},
  volume    = {81},
  doi       = {10.1103/PhysRevLett.81.4785},
  url       = {https://doi.org/10.1103/PhysRevLett.81.4785},
}

@Article{weinorman,
  author    = {Wei, J. and Norman, E.},
  journal   = {J. Math. Phys.},
  title     = {Lie Algebraic Solution of Linear Differential Equations},
  year      = {1963},
  month     = apr,
  number    = {4},
  pages     = {575--581},
  volume    = {4},
  doi       = {10.1063/1.1703993},
  publisher = {AIP Publishing},
  url       = {https://doi.org/10.1063/1.1703993},
}

@Article{qvarfort,
  author    = {Qvarfort, Sofia and Pikovski, Igor},
  journal   = {PRX Quantum},
  title     = {Solving Quantum Dynamics with a {L}ie-Algebra Decoupling Method},
  year      = {2025},
  month     = jan,
  number    = {1},
  pages     = {010201},
  volume    = {6},
  doi       = {10.1103/PRXQuantum.6.010201},
  publisher = {American Physical Society},
  url       = {https://doi.org/10.1103/PRXQuantum.6.010201},
}

@Article{magnus,
  author    = {Magnus, Wilhelm},
  journal   = {Commun. Pure Appl. Math.},
  title     = {On the exponential solution of differential equations for a linear operator},
  year      = {1954},
  issn      = {1097-0312},
  month     = nov,
  number    = {4},
  pages     = {649--673},
  volume    = {7},
  doi       = {10.1002/cpa.3160070404},
  publisher = {Wiley},
  url       = {https://doi.org/10.1002/cpa.3160070404},
}

@Article{wilcox,
  author    = {Wilcox, R. M.},
  journal   = {J. Math. Phys.},
  title     = {Exponential Operators and Parameter Differentiation in Quantum Physics},
  year      = {1967},
  issn      = {1089-7658},
  month     = apr,
  number    = {4},
  pages     = {962--982},
  volume    = {8},
  doi       = {10.1063/1.1705306},
  publisher = {AIP Publishing},
  url       = {https://doi.org/10.1063/1.1705306},
}

@Article{blanes2009,
  author    = {Blanes, S. and Casas, F. and Oteo, J. A. and Ros, J.},
  journal   = {Phys. Rep.},
  title     = {The {Magnus} expansion and some of its applications},
  year      = {2009},
  issn      = {0370-1573},
  month     = jan,
  number    = {5--6},
  pages     = {151--238},
  volume    = {470},
  doi       = {10.1016/j.physrep.2008.11.001},
  publisher = {Elsevier BV},
  url       = {https://doi.org/10.1016/j.physrep.2008.11.001},
}

@Article{moan2006,
  author    = {Moan, Per Christian and Niesen, Jitse},
  journal   = {Found. Comput. Math.},
  title     = {Convergence of the {Magnus} Series},
  year      = {2008},
  issn      = {1615-3383},
  month     = jun,
  number    = {3},
  pages     = {291--301},
  volume    = {8},
  doi       = {10.1007/s10208-007-9010-0},
  publisher = {Springer Science and Business Media LLC},
  url       = {https://doi.org/10.1007/s10208-007-9010-0},
}

@Article{fer,
  author    = {Fer, Francis},
  journal   = {Bull. Cl. Sci., Acad. R. Belg.},
  title     = {R{\'{e}}solution de l'{\'{e}}quation matricielle $d{U}/dt = p{U}$ par produit infini d'exponentielles matricielles},
  year      = {1958},
  issn      = {0001-4141},
  number    = {1},
  pages     = {818--829},
  volume    = {44},
  doi       = {10.3406/barb.1958.68918},
  publisher = {PERSEE Program},
  url       = {https://doi.org/10.3406/barb.1958.68918},
}

@Article{rabi,
  author    = {Rabi, I. I.},
  journal   = {Phys. Rev.},
  title     = {Space Quantization in a Gyrating Magnetic Field},
  year      = {1937},
  issn      = {0031-899X},
  month     = apr,
  number    = {8},
  pages     = {652--654},
  volume    = {51},
  doi       = {10.1103/physrev.51.652},
  publisher = {American Physical Society (APS)},
  url       = {https://doi.org/10.1103/PhysRev.51.652},
}

@Article{bloch,
  author    = {Bloch, F.},
  journal   = {Phys. Rev.},
  title     = {Nuclear Induction},
  year      = {1946},
  issn      = {0031-899X},
  month     = oct,
  number    = {7--8},
  pages     = {460--474},
  volume    = {70},
  doi       = {10.1103/physrev.70.460},
  publisher = {American Physical Society (APS)},
  url       = {https://doi.org/10.1103/PhysRev.70.460},
}

@Article{feynman,
  author    = {Feynman, Richard P. and Vernon, Frank L. and Hellwarth, Robert W.},
  journal   = {J. Appl. Phys.},
  title     = {Geometrical Representation of the {S}chr{\"{o}}dinger Equation for Solving Maser Problems},
  year      = {1957},
  issn      = {1089-7550},
  month     = jan,
  number    = {1},
  pages     = {49--52},
  volume    = {28},
  doi       = {10.1063/1.1722572},
  publisher = {AIP Publishing},
  url       = {https://doi.org/10.1063/1.1722572},
}

@Book{sakurai,
  author    = {Sakurai, J. J. and Napolitano, Jim},
  publisher = {Cambridge University Press},
  title     = {Modern Quantum Mechanics},
  year      = {2020},
  address   = {Cambridge},
  edition   = {3rd},
  isbn      = {9781108473224},
  month     = sep,
  doi       = {10.1017/9781108587280},
  url       = {https://doi.org/10.1017/9781108587280},
}

@Article{Scherer1995,
  author        = {Scherer, Wolfgang},
  journal       = {Phys. Rev. Lett.},
  title         = {Superconvergent Perturbation Method in Quantum Mechanics},
  year          = {1995},
  month         = feb,
  pages         = {1495--1499},
  volume        = {74},
  doi           = {10.1103/PhysRevLett.74.1495},
  numpages      = {0},
  publisher     = {American Physical Society},
  url           = {https://doi.org/10.1103/PhysRevLett.74.1495},
}

@Article{Scherer1998,
  author    = {Scherer, Wolfgang},
  journal   = {J. Math. Phys.},
  title     = {Quantum averaging. {III}. {T}ime-dependent systems},
  year      = {1998},
  issn      = {1089-7658},
  month     = may,
  number    = {5},
  pages     = {2597--2634},
  volume    = {39},
  doi       = {10.1063/1.532411},
  publisher = {AIP Publishing},
  url       = {https://doi.org/10.1063/1.532411},
}

@Article{Vandersypen2005,
  author    = {Vandersypen, Lieven M. K. and Chuang, Isaac L.},
  journal   = {Rev. Mod. Phys.},
  title     = {{NMR} techniques for quantum control and computation},
  year      = {2005},
  month     = jan,
  number    = {4},
  pages     = {1037--1069},
  volume    = {76},
  doi       = {10.1103/RevModPhys.76.1037},
  publisher = {American Physical Society},
  url       = {https://doi.org/10.1103/RevModPhys.76.1037},
}

@Article{Barnes2012,
  author    = {Barnes, Edwin and Das Sarma, Sankar},
  journal   = {Phys. Rev. Lett.},
  title     = {Analytically Solvable Driven Time-Dependent Two-Level Quantum Systems},
  year      = {2012},
  month     = aug,
  number    = {6},
  pages     = {060401},
  volume    = {109},
  doi       = {10.1103/PhysRevLett.109.060401},
  publisher = {American Physical Society},
  url       = {https://doi.org/10.1103/PhysRevLett.109.060401},
}

@Article{Messina2014,
  author    = {Messina, A. and Nakazato, H.},
  journal   = {J. Phys. A: Math. Theor.},
  title     = {Analytically solvable {H}amiltonians for quantum two-level systems and their dynamics},
  year      = {2014},
  month     = oct,
  number    = {44},
  pages     = {445302},
  volume    = {47},
  doi       = {10.1088/1751-8113/47/44/445302},
  url       = {https://doi.org/10.1088/1751-8113/47/44/445302},
}

@Article{klarsfeld1989,
  author  = {Klarsfeld, S. and Oteo, J. A.},
  journal = {Phys. Rev. A},
  title   = {Recursive generation of higher-order terms in the {M}agnus expansion},
  year    = {1989},
  volume  = {39}, number = {7}, pages = {3270--3273},
  doi     = {10.1103/PhysRevA.39.3270},
  url     = {https://doi.org/10.1103/PhysRevA.39.3270},
}

@Article{blanes1998,
  author  = {Blanes, S. and Casas, F. and Oteo, J. A. and Ros, J.},
  journal = {J. Phys. A: Math. Gen.},
  title   = {{M}agnus and {F}er expansions for matrix differential equations: the convergence problem},
  year    = {1998},
  volume  = {31}, number = {1}, pages = {259--268},
  doi     = {10.1088/0305-4470/31/1/023},
  url     = {https://doi.org/10.1088/0305-4470/31/1/023},
}

@Article{casas2007,
  author  = {Casas, Fernando},
  journal = {J. Phys. A: Math. Theor.},
  title   = {Sufficient conditions for the convergence of the {M}agnus expansion},
  year    = {2007},
  volume  = {40}, number = {50}, pages = {15001--15017},
  doi     = {10.1088/1751-8113/40/50/006},
  url     = {https://doi.org/10.1088/1751-8113/40/50/006},
}

@Article{iserles1999,
  author  = {Iserles, A. and N{\o}rsett, S. P.},
  journal = {Philos. Trans. R. Soc. Lond. A},
  title   = {On the solution of linear differential equations in {L}ie groups},
  year    = {1999},
  volume  = {357}, number = {1754}, pages = {983--1019},
  doi     = {10.1098/rsta.1999.0362},
  url     = {https://doi.org/10.1098/rsta.1999.0362},
}

@Article{iserles2000,
  author  = {Iserles, Arieh and Munthe-Kaas, Hans Z. and N{\o}rsett, Syvert P. and Zanna, Antonella},
  journal = {Acta Numer.},
  title   = {{L}ie-group methods},
  year    = {2000},
  volume  = {9}, pages = {215--365},
  doi     = {10.1017/S0962492900002154},
  url     = {https://doi.org/10.1017/S0962492900002154},
}

@Article{zanna1999,
  author  = {Zanna, A.},
  journal = {SIAM J. Numer. Anal.},
  title   = {Collocation and relaxed collocation for the {F}er and the {M}agnus expansions},
  year    = {1999},
  volume  = {36}, number = {4}, pages = {1145--1182},
  doi     = {10.1137/S0036142997326616},
  url     = {https://doi.org/10.1137/S0036142997326616},
}

@Article{weinorman1964,
  author  = {Wei, J. and Norman, E.},
  journal = {Proc. Am. Math. Soc.},
  title   = {On global representations of the solutions of linear differential equations as a product of exponentials},
  year    = {1964},
  volume  = {15}, number = {2}, pages = {327--334},
  doi     = {10.1090/S0002-9939-1964-0160009-0},
  url     = {https://doi.org/10.1090/S0002-9939-1964-0160009-0},
}

@Article{guerrero2020,
  author  = {Guerrero, Julio and Berrondo, Manuel},
  journal = {J. Math. Phys.},
  title   = {Semiclassical interpretation of {W}ei--{N}orman factorization for {SU}(1,1) and its related integral transforms},
  year    = {2020},
  volume  = {61}, pages = {082107},
  doi     = {10.1063/1.5143586},
  url     = {https://doi.org/10.1063/1.5143586},
}

@Article{charzynski2013,
  author  = {Charzy{\'n}ski, Szymon and Ku{\'s}, Marek},
  journal = {J. Phys. A: Math. Theor.},
  title   = {{W}ei--{N}orman equations for a unitary evolution},
  year    = {2013},
  volume  = {46}, number = {26}, pages = {265208},
  doi     = {10.1088/1751-8113/46/26/265208},
  url     = {https://doi.org/10.1088/1751-8113/46/26/265208},
}

@Article{altafini2005,
  author  = {Altafini, Claudio},
  journal = {Syst. Control Lett.},
  title   = {Explicit {W}ei--{N}orman formulae for matrix {L}ie groups via {P}utzer's method},
  year    = {2005},
  volume  = {54}, number = {11}, pages = {1121--1130},
  doi     = {10.1016/j.sysconle.2005.03.006},
  url     = {https://doi.org/10.1016/j.sysconle.2005.03.006},
}

@Article{lewis1969,
  author  = {Lewis, H. R. and Riesenfeld, W. B.},
  journal = {J. Math. Phys.},
  title   = {An exact quantum theory of the time-dependent harmonic oscillator and of a charged particle in a time-dependent electromagnetic field},
  year    = {1969},
  volume  = {10}, number = {8}, pages = {1458--1473},
  doi     = {10.1063/1.1664991},
  url     = {https://doi.org/10.1063/1.1664991},
}

@Article{lloyd1996,
  author = {Lloyd, Seth}, journal = {Science},
  title = {Universal Quantum Simulators},
  year = {1996}, volume = {273}, number = {5278}, pages = {1073--1078},
  doi = {10.1126/science.273.5278.1073}, url = {https://doi.org/10.1126/science.273.5278.1073},
}

@Article{childs2021,
  author = {Childs, Andrew M. and Su, Yuan and Tran, Minh C. and Wiebe, Nathan and Zhu, Shuchen},
  journal = {Phys. Rev. X}, title = {Theory of {T}rotter Error with Commutator Scaling},
  year = {2021}, volume = {11}, number = {1}, pages = {011020},
  doi = {10.1103/PhysRevX.11.011020}, url = {https://doi.org/10.1103/PhysRevX.11.011020},
}

@Article{viola1999,
  author = {Viola, Lorenza and Knill, Emanuel and Lloyd, Seth}, journal = {Phys. Rev. Lett.},
  title = {Dynamical Decoupling of Open Quantum Systems},
  year = {1999}, volume = {82}, number = {12}, pages = {2417--2421},
  doi = {10.1103/PhysRevLett.82.2417}, url = {https://doi.org/10.1103/PhysRevLett.82.2417},
}

@Article{haeberlen1968,
  author = {Haeberlen, U. and Waugh, J. S.}, journal = {Phys. Rev.},
  title = {Coherent Averaging Effects in Magnetic Resonance},
  year = {1968}, volume = {175}, number = {2}, pages = {453--467},
  doi = {10.1103/PhysRev.175.453}, url = {https://doi.org/10.1103/PhysRev.175.453},
}

@Article{mananga2011,
  author = {Mananga, Eug\`{e}ne S. and Charpentier, Thibault}, journal = {J. Chem. Phys.},
  title = {Introduction of the {F}loquet--{M}agnus Expansion in Solid-State Nuclear Magnetic Resonance Spectroscopy},
  year = {2011}, volume = {135}, number = {4}, pages = {044109},
  doi = {10.1063/1.3610943}, url = {https://doi.org/10.1063/1.3610943},
}

@Article{takegoshi2015,
  author = {Takegoshi, K. and Miyazawa, Norihiro and Sharma, Kshama and Madhu, P. K.}, journal = {J. Chem. Phys.},
  title = {Comparison among {M}agnus/{F}loquet/{F}er Expansion Schemes in Solid-State {NMR}},
  year = {2015}, volume = {142}, number = {13}, pages = {134201},
  doi = {10.1063/1.4916324}, url = {https://doi.org/10.1063/1.4916324},
}

@Article{koch2022,
  author = {Koch, Christiane P. and Boscain, Ugo and Calarco, Tommaso and Dirr, Gunther and Filipp, Stefan and Glaser, Steffen J. and Kosloff, Ronnie and Montangero, Simone and Schulte-Herbr\"{u}ggen, Thomas and Sugny, Dominique and Wilhelm, Frank K.}, journal = {EPJ Quantum Technol.},
  title = {Quantum Optimal Control in Quantum Technologies. {S}trategic Report on Current Status, Visions and Goals for Research in {E}urope},
  year = {2022}, volume = {9}, pages = {19},
  doi = {10.1140/epjqt/s40507-022-00138-x}, url = {https://doi.org/10.1140/epjqt/s40507-022-00138-x},
}

@Article{pinney1950,
  author = {Pinney, Edmund}, journal = {Proc. Am. Math. Soc.},
  title = {The Nonlinear Differential Equation $y'' + p(x) y + c y^{-3} = 0$},
  year = {1950}, volume = {1}, number = {5}, pages = {681},
  doi = {10.1090/S0002-9939-1950-0037979-4}, url = {https://doi.org/10.1090/S0002-9939-1950-0037979-4},
}

@Article{Zener1932,
  author  = {Zener, C.},
  journal = {Proc. R. Soc. Lond. A},
  title   = {Non-Adiabatic Crossing of Energy Levels},
  year    = {1932},
  volume  = {137}, number = {833}, pages = {696--702},
  doi     = {10.1098/rspa.1932.0165},
  url     = {https://doi.org/10.1098/rspa.1932.0165},
}

@Article{GueryOdelin2019,
  author  = {Gu{\'e}ry-Odelin, David and Ruschhaupt, Andreas and Kiely, Anthony and Torrontegui, Erik and Mart{\'i}nez-Garaot, Sara and Muga, J. G.},
  journal = {Rev. Mod. Phys.},
  title   = {Shortcuts to adiabaticity: Concepts, methods, and applications},
  year    = {2019},
  volume  = {91}, number = {4}, pages = {045001},
  doi     = {10.1103/RevModPhys.91.045001},
  url     = {https://doi.org/10.1103/RevModPhys.91.045001},
}

@Misc{bagnasacco2026fer,
  author        = {Bagnasacco, Lorenzo and Giovannetti, Vittorio},
  title         = {Improved convergence radius of the {F}er expansion for {H}ermitian generators},
  year          = {2026},
  eprint        = {2609.28302},
  archivePrefix = {arXiv},
}

\end{document}